\documentclass{article}
\usepackage[margin=1in]{geometry}
\usepackage{setspace}
\usepackage{amsmath}
\usepackage{graphicx}
\usepackage[utf8]{inputenc}
\usepackage[table]{xcolor}
\usepackage{cite}
\graphicspath{ {Figures/} }
\usepackage{multicol}
\usepackage{array}
\usepackage{tabu}
\usepackage{ulem}
\date{}
\date{\vspace{-10ex}}
\begin{document}
\title{Coherent Atomic Orbital Polarization Probes the Geometric Phase in Photodissociation of Polyatomic Molecules}
\maketitle

\renewcommand{\thefootnote}{\fnsymbol{footnote}}

\author{Chaya Weeraratna,$^1$}
\author{Arthur G. Suits,$^1$,}\footnote{email: suitsa@missouri.edu}
\author{Oleg S. Vasyutinskii$^2$,}\footnote{email: osv@pms.ioffe.ru}\\
$^1$Department of Chemistry, University of Missouri, Columbia MO 65211 USA\\
$^2$Ioffe Institute, Russian Academy of Sciences, St. Petersburg 194021 Russia \\

\section*{Abstract} 

Quantum interference between multiple pathways in molecular photodissociation often results in angular momentum polarization of atomic products and this can give deep insight into fundamental physical processes. For dissociation of diatomic molecules the resulting orbital polarization is fully understood and consistent with quantum mechanical theory. For polyatomic molecules, however, coherent photofragment orbital polarization is frequently observed but so far has eluded theoretical explanation, and physical insight is lacking. Here we present a model of these effects for ozone photodissociation that reveals the importance of a novel manifestation of the geometric phase. We show this geometric phase effect permits the existence of coherent polarization in cases where it would otherwise vanish, and cancels it in some cases where it might otherwise exist. The model accounts for measurements in ozone that have hitherto defied explanation, and represents a step toward a deeper understanding of coherent electronic excitation in polyatomic molecules and a new  role of the geometric phase.

\newpage

When two electronic states are simultaneously excited in a photodissociation process, quantum interference can occur and this may be manifested in the asymptotic region as orbital polarization of the atomic products.~\cite{vigue1981polarization,siebbeles1994vector,rakitzis1998photofragment,bracker1998observation,bracker1999imaging,korovin2000observation} This interference allows one to determine, directly from experiment, fundamental physical values that appear in the complete quantum mechanical theory but are otherwise inaccessible.  As shown by Picheyev el al.~\cite{picheyev1997ground} and by Rakitzis and Zare~\cite{Rakitzis&Zare1999} in general these values can be expressed in terms of a set of \emph{coherent} anisotropy parameters that contain information about the photodissociation dynamics. This quantum interference always exists in a typical  case when the dissociative transition at a fixed photon energy can prepare more than one excited state, and the molecule can break apart on several potential energy surfaces leading to the same product  simultaneously~\cite{siebbeles1994vector,rakitzis1998photofragment}.

\begin{figure}[h]
\centering
\includegraphics[width=0.8\textwidth]{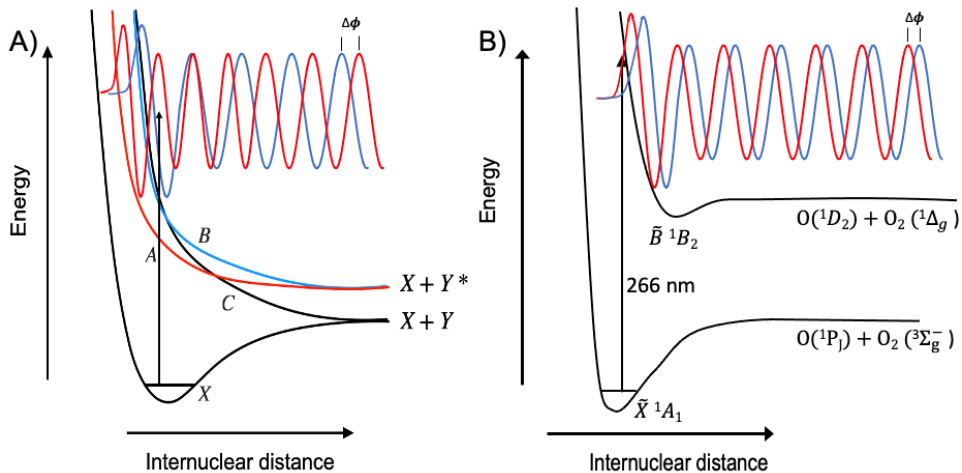}
\caption{A) Scheme describing an energy-sensitive phase shift controlling the production of photofragment angular momentum orientation in a general photodissociation event. B) Schematic illustration of the relevant potentials in ozone illustrating the geometric phase effect.}
\label{fig:diatomic}
\end{figure}

One of these values is a quantum mechanical phase shift $\Delta\varphi$  which can be associated with the phase shift between de Broglie matter waves following these multiple dissociating pathways.~\cite{rakitzis1998photofragment,BalintKurti2002} In their seminal paper, Rakitzis et al.\cite{rakitzis1998photofragment} have demonstrated that orientation (helicity) of the photofragment electron angular momentum $\mathbf{j}$ caused by quantum interference between multiple pathways in photodissociation of a diatomic molecule with linearly polarized light is proportional to $\sin\Delta\varphi$. In the experiment of Rakitzis et al., the absorption of a single photon in the ICl diatomic molecule promoted simultaneously parallel and perpendicular transitions from the $^1\Sigma^+$ ground state to $^3\Pi_{0^+}$ and $^3\Pi_1$ excited electronic states, respectively.~\cite{rakitzis1998photofragment} This so-named \emph{coherent} parallel and perpendicular absorption of a linearly polarized photon caused electron charge oscillations parallel and perpendicular to the internuclear axis with the frequency difference directly related to the energy difference between the interfering dissociating pathways, schematically shown in Fig.\ref{fig:diatomic}A. As the nuclei separated, the frequency difference of oscillation changed and at large nuclear separations vanished. The resulting scattering wave function accumulated a net phase shift $\Delta\varphi$ that depended strongly on the excitation photon energy. This in turn resulted in observed oscillation in the helicity  as a function of the wavelength of the photolysis light used in the experiment~\cite{rakitzis1998photofragment}. The analogous scheme for the case of ozone illustrating the role of the geometric phase is given in Fig.\ref{fig:diatomic}B and discussed below.

Experimental studies of photofragment electronic angular momentum helicity, produced in photodissociation of diatomic molecules with circularly polarized light~\cite{korovin2000observation,Alexander2000}, typically showed strong effects as a result of both incoherent and coherent dissociation mechanisms. According to the theoretical analysis of Balint-Kurti et al.~\cite{BalintKurti2002} in the case of coherent dissociation with circularly polarized light, the photofragment angular momentum helicity is proportional to $\cos\Delta\varphi$. All the above physical mechanisms for the case of diatomic molecules  are now well understood and perfectly agree with predictions of full quantum mechanical theory ~\cite{Suits&Vasyutinskii2008}.

The situation for polyatomic molecules is very different, however. Although studies of angular momentum polarization of atomic products in photodissociation of polyatomics have been carried out intensively during recent years and have produced many remarkable and challenging results~\cite{mo1996simultaneous,Teule2000(N2O),dylewski2001energy,rakitzis2001complete,lee20051,Brom2005,brouard2006(O3),Denzer2006,brouard2007(O3),
Chen2007(ICN),brouard2007a(OCS),brouard2007b(OCS),lipciuc2011slice,Corrales2016}, the understanding of these effects is poor and incomplete. Most of the experiments have shown large polarization effects, and sets of the orientation and alignment parameters have been determined in many studies. Although each of the parameters  labels incoherent/coherent types of excitation and parallel/perpendicular excitation channels in arbitrary diatomic or polyatomic molecules~\cite{kuznetsov2007photofragment,shternin2008,Suits&Vasyutinskii2008,shternin2012}, the latter distinction is strictly defined only for diatomic molecules, and in the generalization to polyatomic case, profound differences arise that have precluded the development of a general understanding of these phenomena.

The concept of the Geometric Phase (GP) in quantum systems has been known for decades and has become an essential aspect of understanding and interpreting many important physical phenomena.~\cite{berry1990anticipations,xiao2010berry,Vanderbilt2018}. Application of the GP to molecular systems was first performed by Herzberg and Longuet-Higgins~\cite{Herzberg63} who considered adiabatic dynamics of a doubly degenerate electronic state in the vicinity of a conical intersection (CI) and showed that the electronic wavefunction changes sign upon encircling the CI. Mead and Truhlar first considered the implications for reaction on a single potential surface involving a CI for identical nuclei~\cite{mead1979determination}, anticipating the great interest in implications for the hydrogen exchange reaction which have recently led to the first definitive experimental evidence of the effect.~\cite{yuan2020observation}   Nix et al~\cite{nix2008observation} reported evidence for the importance of the GP effect in photodissociation of phenol, and the important role of nonadiabatic tunneling mediating the effect has been revealed by Xie, Yarkony and Guo  in several papers. \cite{xie2016nonadiabatic,xie2019insights} Recent work has shown that the GP effect can profoundly modulate reactivity in ultracold reactions as well.~\cite{kendrick2015geometric} In all of these molecular systems discussed to-date the GP effect arises owing to the presence of the CI as initially described by Herzberg and Longuet-Higgins.   In highly celebrated work, Berry broadly generalized the GP approach for all adiabatic processes\cite{berry1990anticipations} by considering the adiabatic evolution of a quantum system through a loop in a parameter space.

Here we show that a key aspect of understanding coherent orbital polarization in dissociation of polyatomic molecules is a recognition that this polarization directly manifests geometric phase (GP) effects, and a model we develop incorporating these effects qualitatively accounts for fundamental aspects of ozone photodissocation seen in many experiments. We suggest that although the coherent anisotropy parameters always relate to a certain quantum mechanical phase shift, the nature of the phase shifts in diatomic and polyatomic cases can be completely different.  In contrast to the widely recognized role of GP effects in the vicinity of CIs, this is a fundamentally different manifestation of the GP that appears in photodissociation of planar polyatomic molecules, and it  may either extinguish or give rise to coherent orbital polarization of the atomic photofragments. The effect arises owing to coherent photoexcitation of degenerate vibrational states (Herzberg-Teller interactions) rather than from nuclear motion involving a CI. Owing to the coherent excitation, the electronic wavefunction creates a loop resulting in a phase difference which has a pure geometric origin. The orbital polarization in the atomic photofragments that results from this excitation directly reports on this phase difference and the GP effect.  In some cases a CI can also contribute to the wavefunction phase shift, and the coherent parameters will also be sensitive to this. We develop this model based on experiments~\cite{lee20051,Denzer2006,weeraratna2019photodissociation} on ozone photolysis via the Hartley band involving absorption to, and diabatic dissociation on the $\tilde{B}\,^1B_2 (3\, ^1A_1)$ potential energy surface and resulting in orbital polarization of the atomic O($^1D_2$) fragments. The model is supported by a full quantum mechanical theory developed by the authors~\cite{Vasyutinskii2019}. The paper is organized as follows: we first summarize the UV photodissociation of ozone and the existing experimental results for coherent orbital polarization in this system. We then outline the role of the GP in coherent angular momentum polarization. This is followed by presentation of our model of ozone dissociation by linearly polarized light and the consequences of the GP for the associated orientation, then the complementary picture for dissociation by circularly polarized light and the associated coherent alignment. Additional details concerning the theory are presented in Supplementary Information (SI).

\section*{Results and discussion}
\subsection*{Coherent Orbital Polarization in Ozone Photolysis}
The ozone absorption spectrum in the Hartley band extends from 200 nm - 310 nm peaking at 254 nm. Under excitation at 266 nm the molecule is excited almost exclusively to the \~B electronic state (see Fig.~\ref{fig:diatomic}B), however nonadiabatic transitions to \~A, \~R and the ground state \~X are important for complete analysis~\cite{baloitcha2005theory,schinke2010photodissociation}. Upon excitation the dissociation happens via the following spin-conserving channels:
\begin{eqnarray}
\label{eq:path1}
{\text O}_3(~X^1A_1) + \textit{hv} \rightarrow {\text O}_2(a^1\Delta_g) + {\text O}(^1D_2)  \\
\label{eq:path2}
{\text O}_3(~X^1A_1) + \textit{hv} \rightarrow {\text O}_2(X^3\Sigma^{-}_{g}) + {\text O}(^3P_J)
\end{eqnarray}


\noindent The major product channel ("singlet channel") is the pathway (\ref{eq:path1}) accounting for 92\% of the oxygen atomic fragments.  The remaining 8\% are produced via the "triplet channel" (\ref{eq:path2}) due to a nonadiabatic transition between \~B and \~R states~\cite{baloitcha2005theory,schinke2010photodissociation}. In the absence of vibration in the ground state ozone belongs to the $C_{2v}$ point group, where \~X, \~A and \~R states have $^1A_1$ symmetry and the  \~B state has $^1B_2$ symmetry. When the molecule dissociates the $C_{2v}$ symmetry is lost, then all the states have $^1A^\prime$ symmetry.

\begin{figure}[h]
    \centering
        \includegraphics[width=0.8\textwidth]{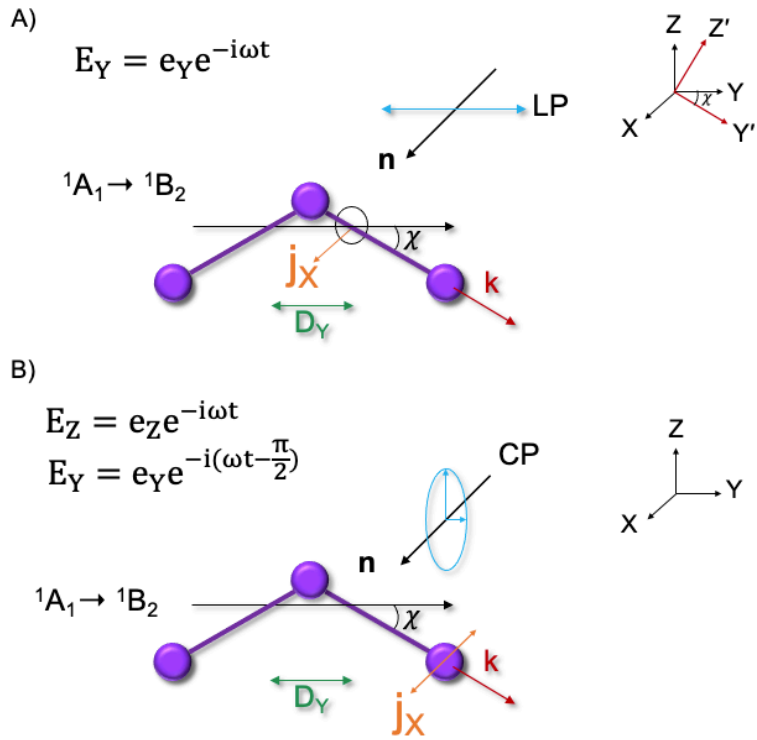}
    \caption{ Models of orbital orientation and alignment in ozone photolysis via the $\tilde B(^1B_2) \leftarrow \tilde X(^1A_1) $ transition by linearly polarized (LP) and circularly polarized (CP) light.
    The molecule is in the YZ plane where Z is the symmetry axis,  $\mathbf{n}$ is the laser propagation direction, $\mathbf{k}$ is the fragment recoil direction, and D$_Y$ is the transition dipole moment.
    A) LP light. A circle with an arrow denotes the direction of electron cloud oscillation producing the photofragment angular momentum $\mathbf{j}_x$ perpendicular to the molecular plane.
    B) CP light. $\mathbf{j}_x$ is the photofragment angular momentum perpendicular to the molecular plane. }
    \label{fig:O3Model}
\end{figure}

According to the selection rules for $C_{2v}$ symmetry~\cite{Herzberg91} the major dipole transition $\tilde{B}^1B_2 \leftarrow \tilde{X}^1A_1$  is induced by a single in-plane body frame electronic dipole moment component $D_Y$ that is perpendicular to the ozone symmetry axis, see Fig.~\ref{fig:O3Model}.  Optical excitation in ozone occurs into a single electronic potential energy surface (PES)  \~B and the condition of  parallel/perpendicular coherent excitation of two excited states at  first glance cannot be fulfilled. Therefore, within this simplified model, no coherent optical polarization can be produced in ozone photolysis neither with linearly, nor circularly polarized light.

\begin{figure}[h]
    \centering
        \includegraphics[width=0.5\textwidth]{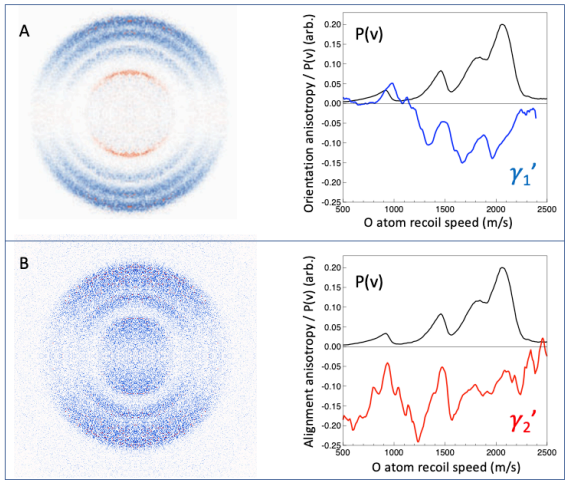}
    \caption{ DC sliced velocity map difference images for O($^1$D) from ozone photolysis at 266 nm showing angular momentum polarization. Plots of the recoil speed dependence of the indicated anisotropy parameters are given on the right. A) Orientation produced by LP photolysis light. Adapted from reference 16. B) Alignment produced by CP light. Adapted from reference 39. The overall speed distribution, P(v), is given in each case with peaks corresponding to vibrational levels 0-3 in the O$_2$ cofragment.   }
    \label{fig:images}
\end{figure}

However, recent sliced imaging experiments on ozone photolysis at 266~nm reproduced in Fig.~\ref{fig:images} demonstrated remarkably high angular momentum orientation~\cite{lee20051} and alignment~\cite{weeraratna2019photodissociation} in the ${\text O}(^1D_2)$ fragments produced in channel (\ref{eq:path1}) with linearly and circularly polarized photolysis light, respectively.  Lee et al~\cite{lee20051} and Denzer et al~\cite{Denzer2006} reported that of the three possible orientation mechanisms in ozone photodissociation, only that arising from the coherent excitation with linearly polarized light (described by the orientation parameter $\gamma_1'$) was nonzero. Moreover, it was demonstrated~\cite{lee20051} that the $\gamma_1'$ parameter depended strongly but in a non-oscillating way on the excitation photon energy, and changed its sign at low fragment recoil velocity (See Fig.~\ref{fig:images}A). Another important result has recently been reported by Weeraratna et al.~\cite{weeraratna2019photodissociation} who observed the orbital alignment of O($^1D_2$) fragments induced by circularly polarized photolysis light. This effect is described by the $\gamma_2^\prime$ anisotropy parameter. These results are shown in Fig.~\ref{fig:images}B. As can be seen, the anisotropy parameter $\gamma_1'$ decreases monotonically with the fragment velocity and changes its sign at the low end, while the parameter $\gamma_2'$ shows only a modest dependence on the fragment velocity and no change of sign. According to the general quantum mechanical nomenclature~\cite{wouters2001imaging,Suits&Vasyutinskii2008} both effects observed are due to coherent parallel/perpendicular photoexcitation. Also, the direction of the fragment angular momentum produced in the body frame in both cases was perpendicular to the recoil direction.

As shown below, the mechanism of coherent excitation can occur within the planar symmetry of ozone and it is not the same in cases of linearly and circularly polarized excitation. In the former case, the effect is mainly because the recoil direction is not collinear with the transition dipole moment direction and the required combination of  parallel/perpendicular transition dipole matrix elements occurs due to re-direction of the dipole moment components onto the molecular bond (recoil) direction. In the latter case the coherent excitation is caused mostly by vibronic interactions involving symmetric and asymmetric vibrational modes in the $\tilde{B}\,{^1B_2}$ excited state~\cite{weeraratna2019photodissociation}. In both cases the angular momentum polarization is proportional to $\sin\Delta\varphi$, where, unlike photolysis of a diatomic molecule, the phase difference $\Delta\varphi$ relates to a GP making a loop in the configuration space along  PES \~B. Possible nonadiabatic interactions is the vicinity of CIs  are in general not necessary for explanation of both angular momentum polarization effects however they can contribute to the phase shift $\Delta\varphi$ and the contribution is different in cases of linearly and circularly polarized excitation.

\subsection*{Coherent Angular Momentum Polarization and the Geometric Phase}
\begin{figure}[h]
\centering
\includegraphics[width=0.6\textwidth]{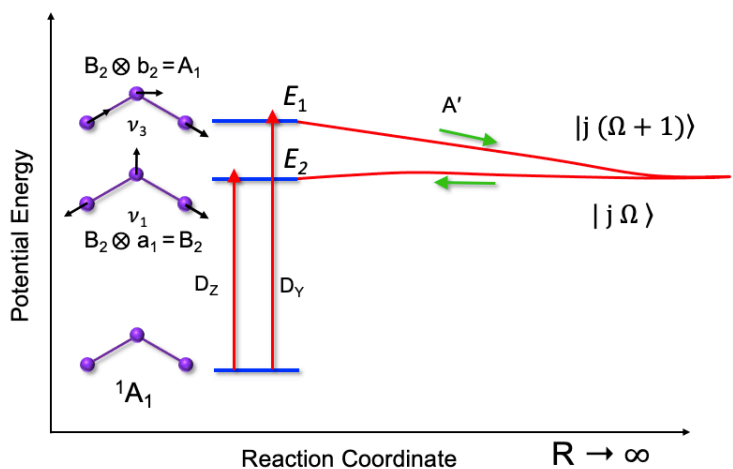}
\caption{The loop in parameter space schematically illustrating the GP mechanism in ozone photolysis.}
\label{fig:loop1}
\end{figure}

For theoretical description of the orbital orientation and alignment produced by ozone photolysis we employ a set of anisotropy transforming coefficients $\mathbf{c}^K_{k_dq_k}$ introduced by Shternin and Vasyutinskii~\cite{shternin2008} which are directly related to anisotropy parameters $\gamma'_1$ and $\gamma'_2$ as presented  in Supplementary Information (SI). It is shown that these coefficients can be written as a sequence of transition matrix elements that include the Y and Z electronic dipole moment components coupling the ground state to the asymptotic scattering wavefunctions with distinct angular momentum projections onto the recoil direction. We show this sequence of transition matrix elements makes a closed loop in parameter space schematically shown in Fig.\ref{fig:loop1}. The energy levels $E_1$ and $E_2$ in Fig.\ref{fig:loop1} are shifted one from the other for clarity: in fact they can be either degenerate, or not depending on which  effect is considered.

This expression for the $\mathbf{c}^K_{k_dq_k}$ coefficients includes a phase which can  be presented in the form:
 \begin{eqnarray}
\label{eq:GP}
\Delta\varphi=-Im\,ln[\mathbf{c}^K_{k_d 1}]
\end{eqnarray}
that is a known general GP definition~\cite{Vanderbilt2018}. That is, a phase that arises from adiabatic traversal of a loop in the system's parameter space.
\color{black}
\color{black}

However, unlike the well-known GP that is widely used for interpretation of molecular dynamics problems~\cite{berry1990anticipations,Herzberg63,nix2008observation} that refers to a closed loop in parameter space in the vicinity of a CI, the GP in eq.(\ref{eq:GP}) (see also eq.(1) in SI)  contains also dipole transition matrix elements connecting the ground and excited molecular electronic states. As shown below, in this case the GP contains not only the wavefunction phases, but also the phases of the dipole moments and can differ from zero even if there is no CI inside the loop.

Note that the dipole moment cartesian components $D_i$, $D_j$ directly relate to the corresponding excitation light polarization components $E_i$, $E_j$ due the scalar-product form of the light-matter interaction operator $V_{rad}$:
\begin{eqnarray}
\label{eq:Vrad}
V_{rad} \propto \sum_i E_i D_i.
\label{eq:C}
\end{eqnarray}

According to eq.(\ref{eq:Vrad}), E$_X$, E$_Y$ and E$_Z$ components of the photolysis light electric vector can only excite D$_X$, D$_Y$ and D$_Z$ components of the transition dipole moment, respectively.  Each of the electric field components produces the oscillations of the molecular electric cloud along the corresponding direction.Therefore, for producing the \emph{orientation} of the orbital angular momentum \textbf{j} along the direction $X$ perpendicular to the plane of the molecule schematically shown in Fig.~\ref{fig:O3Model}A,  the electric cloud oscillations along Z and Y axes need to be excited simultaneously with a phase difference of $\pi/2$ between them. Accordingly, for producing the orbital angular momentum  \emph{alignment} along the same direction schematically shown in Fig.~\ref{fig:O3Model}B, a phase difference between the electric cloud oscillations needs to be zero. As shown below  the photofragment angular momentum polarization in photolysis of ozone with linearly and circularly polarized light observed by Lee et al~\cite{lee20051} and Weeraratna et al~\cite{weeraratna2019photodissociation}, respectively is closely related with the GP concept~\cite{xiao2010berry,Vanderbilt2018}. However, the detailed mechanisms of these two effects are not the same so they are discussed below separately.

\subsubsection*{Photolysis with linearly polarized light}
\label{sec:linear}

A model illustrating orbital angular momentum orientation in ozone photolysis via the $\tilde X(^1A_1) \rightarrow \tilde B(^1B_2)$ transition by linearly polarized light is shown in Fig.~\ref{fig:O3Model}A. As only the $D_y$ component of the transition dipole moment differs from zero, $i=j=$Y in eq.(\ref{eq:C}) and the incident light polarized along Y results in electron cloud oscillations in the same direction. However, these oscillations do not initially produce any coherence.

The key point of the model described is that the transition dipole moment $D_Y$ is not parallel to the breaking axis O-O. The breaking axis is denoted as Z$^\prime$ axis of the molecular frame X$^\prime$Y$^\prime$Z$^\prime$ in Fig.~\ref{fig:O3Model}A that is obtained by rotation of the molecular frame XYZ on the angle ($\chi + \pi/2$) around axis X. Therefore, the electric field oscillations along Y induce simultaneous electron cloud excitations along the breaking O-O bond Z$^\prime$ and along the perpendicular in-plane direction Y$^\prime$. These oscillations have a phase shift because of the anisotropy of the electric field inside the molecule. Assuming that the electron cloud oscillation amplitude along Z$^\prime$ follows the function $\cos(\omega t)$, the oscillation along perpendicular direction Y$^\prime$ can in general be described by the function $\cos(\omega t-\Delta\varphi)$, where $\Delta\varphi$ is the phase shift. This function can be presented as a sum of in-phase and quadrature components:
\begin{equation}
\label{eq:quadrature}
\cos(\omega t-\Delta\varphi)=  \sin(\omega t)\sin\Delta\varphi+\cos(\omega t)\cos\Delta\varphi.
\end{equation}

The first term in the \emph{rhs} in eq.~(\ref{eq:quadrature}) (quadrature component) leads to the production of the angular momentum \emph{orientation} $j_x$ shown in Fig.~\ref{fig:O3Model}A and directed perpendicular to the molecular plane. This orientation is described by the anisotropy parameter $\gamma_1'$ observed by Lee et al~\cite{lee20051} and Denzer et al~\cite{Denzer2006}. The second term (in-phase component) leads to the production of the angular momentum \emph{alignment} directed perpendicular to the molecular plane that is described by the anisotropy parameter $\gamma_2$~\cite{Suits&Vasyutinskii2008}. The molecular frame dipole moment components $D_{Y'}$ and $D_{Z'}$ in Fig.~\ref{fig:O3Model}A and in eq.~(\ref{eq:C}) differ from zero leading to the required coherent parallel/perpendicular excitation. This coherent excitation occurs on a single PES and was referred in ref.\cite{lee20051} as "static coherence". The  dipole moment components $D_{Y'}$ and $D_{Z'}$ can be readily transformed to the $X,Y,Z$ frame:
     \begin{eqnarray}
\label{eq:Y'}
D_{Z'}&=&D_Y\cos\chi,\\
D_{Y'}&=&-D_Y\sin\chi=D_Y\cos(\chi-\pi/2).
\label{eq:Z'}
\end{eqnarray}

Equations~(\ref{eq:Y'}) and (\ref{eq:Z'}) show that the sequence of matrix elements in eq.(\ref{eq:C}) along a closed loop results in a constant GP of $\Delta\varphi=\Delta\varphi_{GP}=\pi/2$. At this phase shift value only the first term in eq.(\ref{eq:quadrature}) can differ from zero resulting in fragment orbital angular momentum orientation j$_X$ shown in Fig.~\ref{fig:O3Model}A described by the anisotropy parameter $\gamma_1'$ and experimentally observed in ozone photolysis at 266~nm by Lee et al\cite{lee20051}. In the absence of nonadiabatic interactions this orientation does not depend on the fragment velocity.

This conclusion is changed however, if the photolysis occurs through a conical intersection between the $\tilde B$ and $\tilde A$ states (see Fig. 1 in ref.~\cite{baloitcha2005theory}) where the initial matter flux along $\tilde B$ state is split into two fluxes: along  $\tilde B$ and  $\tilde A$ states due to nonadiabatic transition between them. The nonadiabatic transitions likely occur out of the Frank-Condon area~\cite{baloitcha2005theory}. The nonadiabatic interactions give an additional dynamic phase shift $\Delta\varphi_d$ that depends on the fragment velocity.  Therefore, the total phase shift $\Delta\varphi=\Delta\varphi_{GP} + \Delta\varphi_d$ depends on the fragment velocity. This explains qualitatively the experimental result reported by Lee et al \cite{lee20051} where the angular momentum orientation decreases as a function of the recoil velocity and even change its sign at small velocities.

Another important conclusion resulting from eqs.(\ref{eq:Y'}) and (\ref{eq:Z'}) is that the second (in-phase) term in eq.(\ref{eq:quadrature}) is zero.  Therefore, the alignment parameter $\gamma_2$ describing the contribution from coherent parallel/perpendicular excitation can differ from zero only due to the dynamic phase shift $\Delta\varphi_d$ as the geometric phase shift $\Delta\varphi_{GP}=\pi/2$ is equal to zero in this case. This conclusion is in qualitative agreement with the experimental result of Denzer et al~\cite{Denzer2006} who reported that the contribution to the orbital alignment from the coherent parallel/perpendicular excitation at 298~nm in ozone photolysis was determined to be relatively small.

\subsubsection*{Photolysis with circularly polarized light}
As stated by the general classification based of symmetry arguments~\cite{Suits&Vasyutinskii2008} the photofragment orbital polarization in molecular photolysis via coherent parallel/perpendicular excitation with circularly polarized light described by the anisotropy parameter $\gamma_2'$ can occur if the light beam direction is perpendicular to the recoil axis. A model illustrating orbital angular momentum alignment in ozone photolysis via the $\tilde X(^1A_1) \rightarrow \tilde B(^1B_2)$ transition by circularly polarized light is shown in Fig.~\ref{fig:O3Model}B, where the laser beam direction $\mathbf{n}$ is parallel to axis X and perpendicular to the molecular frame.


As reported by Weeraratna et al~\cite{weeraratna2019photodissociation}, the observed O($^1D_2$) orbital alignment changed its sign when the laser light polarization was switched from right- to left-handed. Therefore, the laser beam helicity labelled by the $\pm\frac{\pi}{2}$ phase shift between the light polarization vectors $E_y$ and $E_z$ in Fig.~\ref{fig:O3Model}B is essential for explanation of the alignment effect. As shown above, within the validity of the Born-Oppenheimer approximation only the $D_y$ component of the molecular dipole moment can be optically excited. During this excitation the information on the phase shift between the light polarization vectors $E_y$ and $E_z$ is completely lost and therefore from the first view $\gamma_2'$ is zero in this conditions.

The above conclusion is changed however if the Born-Oppenheimer approximation fails in the Frank-Condon area. In this case the electronic and the vibrational degree of freedom can interact with each other and the total (vibronic) molecular wavefunctions must be considered. Within the C$_{2v}$ symmetry group, the vibronic wave functions of ozone obey the following symmetries:
\begin{eqnarray}
\label{eq:symmetry1}
^1A^{(tot)}_1 =\, ^1A_1 \otimes\, a_1, \\
^1B^{(tot)}_2 =\, ^1B_2 \otimes\, a_1,
\label{eq:symmetry2} \\
^1A^{(tot)}_1 =\, ^1B_2 \otimes\, b_2 ,
\label{eq:symmetry3}
\end{eqnarray}
where the superscript index $(tot)$  labels the total molecular symmetry, the capital letters in the \emph{rhs} denote electronic irreducible representations, the symbols $\otimes$ denote the irreducible products,  and lowercase letters denote vibrational irreducible representations. Equation~(\ref{eq:symmetry1}) describes the major symmetry of ozone ground state wave function and eqs.~(\ref{eq:symmetry2}), (\ref{eq:symmetry3}) describe possible symmetries of excited state wave functions.

As can be seen in eqs.~(\ref{eq:symmetry2}) and (\ref{eq:symmetry3}), excitation of the symmetric $a_1$ vibrational mode of the $^1B_2$ electronic state reveals the total excited state symmetry $^1B^{(tot)}_2$ while excitation of the  antisymmetric $b_2$ vibrational mode reveals the total excited state symmetry $^1A^{(tot)}_1$.  Therefore, according to the selection rules for $C_{2v}$ symmetry~\cite{Herzberg91} these Herzberg-Teller interactions permit both D$_Y$ and D$_Z$ transition dipole moment components that can be excited simultaneously giving rise to the electron cloud oscillations along Y and Z axes, respectively. Thus, CP photolysis light introduces molecular electron cloud oscillations in two mutually perpendicular directions $Z$ and $Y$ with a phase shift between them.

Assuming that the electron cloud oscillation amplitude along Z follows the function $\cos(\omega t)$, the oscillation along perpendicular direction Y can in general be described by the function $\cos(\omega t-\Delta\varphi-\pi/2)$, where $\Delta\varphi$ is the GP shift and $-\pi/2$ is the initial phase shift between the light polarization vectors $E_y$ and $E_z$ in Fig.~\ref{fig:O3Model}B. As in eq.(\ref{eq:quadrature}) this function can be presented as a sum of in-phase and quadrature components:
\begin{equation}
\label{eq:inphase}
\cos(\omega t-\Delta\varphi-\pi/2)=\sin(\omega t-\Delta\varphi) =  \sin(\omega t)\cos\Delta\varphi-\cos(\omega t)\sin\Delta\varphi.
\end{equation}

The first term in the \emph{rhs} in eq.~(\ref{eq:inphase}) (quadrature component) leads to the production of the angular momentum \emph{orientation} shown in Fig.~\ref{fig:O3Model}A and directed perpendicular to the molecular plane. This orientation is described by the anisotropy parameter $\gamma_1$ that was found to have practically zero value in experiments of Lee et al~\cite{lee20051} and Denzer et al~\cite{Denzer2006}. The second term (in-phase component) leads to the production of the angular momentum \emph{alignment} directed perpendicular to the molecular plane that is described by the anisotropy parameter $\gamma_2'$. This effect was observed experimentally by Weeraratna et al~\cite{weeraratna2019photodissociation}. The molecular frame dipole moment components $D_{Y'}$ and $D_{Z'}$ in Fig.~\ref{fig:O3Model}A and in eq.~(\ref{eq:C}) in SI differ from zero leading to the required coherent parallel/perpendicular excitation.

In this case the excited energy states $E_1$ and $E_2$ in Fig.\ref{fig:loop1}  are almost degenerate in the Frank-Condon area and the corresponding wave functions $\Psi^{(1)}$ and $\Psi^{(2)}$ obey the $^1A_1$ and $^1B_2$ symmetries, respectively. The molecular excited state total wave function $\Psi$ can in the entire $(R,\chi)$ space be written as a linear superposition of the wave functions  $\Psi^{(1)}$ and $\Psi^{(2)}$:
\begin{eqnarray}
\label{eq:Psi}
\Psi = a_1 \Psi^{(1)} + a_2 \Psi^{(2)},
\end{eqnarray}
where $a_1$ and $a_2$ are expansion coefficients that obey the relationship $a_1^2+a_2^2=1$.

According to the Born-Oppenheimer approximation, the wave functions $\Psi^{(1)}$ and $\Psi^{(2)}$ in Eq.~(\ref{eq:Psi}) can be presented everywhere but in the Franck-Condon area as a product of a molecular electronic wave function and a nuclear scattering function $\eta(R,\chi)$ that is a subject of solution of a set of the close-coupling equations\cite{baloitcha2005theory}.
Assuming that the coefficients  $a_1$ and $a_2$ are real, Eq.~(\ref{eq:Psi}) can be rewritten in the form:
\begin{eqnarray}
\label{eq:Psi1}
\Psi = \cos\varphi\, \Psi^{(1)} + \sin\varphi \,\Psi^{(2)}.
\end{eqnarray}
We can evaluate the phase shift $\Delta\varphi$ in the case of excitation with CP light as a loop for the coherent anisotropy parameter with $q_k=1$ as shown in Eq.~(1) in SI.

According to Eq.~(\ref{eq:Psi1}) in the right hand side of the loop in SI Eq.~(1) the phase shift $\varphi$ is equal to zero ($\Psi=\Psi^{(1)}$), while in the left hand side of the loop it is equal to $\varphi=\pi/2$ ($\Psi=\Psi^{(2)}$). Therefore, the phase difference $\Delta\varphi$ is a pure "geometric phase" that is equal to $\Delta\varphi=\pi/2$. The loop occurs in the parametric two-dimensional $R, \chi$ space, along the $A'$ PES where $R$ and $\chi$ are Jacobi coordinates, as shown in Fig.~\ref{fig:loop1}. As mentioned in discussion under Eq.~(\ref{eq:inphase}) in this case the anisotropy transforming coefficient $\mathbf{c}^2_{1 1}=i\,V_2(j)\sqrt{15/2}\,\gamma_2'$ that has a pure imaginary value can differ from zero  while the coefficient $\mathbf{c}^1_{1 1}= (3/2)\sqrt6\,\gamma_1$ has zero value.

The above mechanism does not show fragment velocity dependence of the alignment produced as  both channels photolysis occur along the same $A'$ PES which agrees well with the results of recent experiments\cite{lee20051,weeraratna2019photodissociation}.

Therefore, ozone photolysis under the $\tilde X \rightarrow \tilde B$ excitation with LP light can result in the orbital angular momentum \emph{orientation} of the $O(^1D_2)$ atomic fragments described by the $\gamma'_1$ anisotropy parameter even within the validity of the Born-Oppenheimer approximation, while excitation with CP light can result in orbital angular momentum \emph{alignment} of the $O(^1D_2)$ atomic fragments described by the $\gamma'_2$ anisotropy parameter only under the breakdown of the Born-Oppenheimer approximation. Moreover, unlike photolysis with CP light the $\gamma'_1$ parameter value is sensitive to the nonadiabatic interactions in the molecular excited state and therefore can be used for study of these interactions.

\subsection*{Acknowledgements}
AGS acknowledges the NSF under award number CHE-1955239. OSV is grateful for the Basic foundation for support within the grant No 19-1-1-13-1.

\newpage

\bibliographystyle{unsrt}
\bibliography{Citations}

\end{document}